# Development of input connections in neural cultures

Jordi Soriano, María Rodríguez Martínez, Tsvi Tlusty, and Elisha Moses*

Department of Physics of Complex Systems, Weizmann Institute of Science, Rehovot 76100, Israel

**We introduce an approach for the quantitative assessment of the connectivity in neuronal cultures, based on the statistical mechanics of percolation on a graph. This allows us to monitor the development of the culture and to see the emergence of connectivity in the network. The culture becomes fully connected at a time equivalent to the expected time of birth. The spontaneous bursting activity that characterizes cultures develops in parallel with the connectivity. The average number of inputs per neuron can be quantitatively determined in units of $m_0$, the number of activated inputs needed to excite the neuron. For $m_0 \simeq 15$ we find that hippocampal neurons have on average $\approx$60–120 inputs, whereas cortical neurons have $\approx$75–150, depending on neuronal density. The ratio of excitatory to inhibitory neurons is determined by using the GABA$_A$ antagonist bicuculine. This ratio changes during development and reaches the final value at day 7–8, coinciding with the expected time of the GABA switch. For hippocampal cultures the inhibitory cells comprise $\approx$30% of the neurons in the culture whereas for cortical cultures they are $\approx$20%. Such detailed global information on the connectivity of networks in neuronal cultures is at present inaccessible by any electrophysiological or other technique.**

neural network | network connectivity | inhibition | graph theory | percolation

The formation of the brain is one of the most complicated processes during development. The neural connectivity that initially emerges is organized but imprecise, and further refinement is needed for the accurate formation of the neural circuits. This requires the presence of neural activity (1), first in the form of large-scale spontaneous activity (2), and later driven by sensory experience (3, 4). The connectivity must be flexible enough to allow complex refinement, yet robust enough to sustain synchronous patterns of activity across hundreds of neurons. The formation of connectivity during maturation of the nervous system thus naturally arises as an intriguing issue. Neural cultures have been very useful as model systems to study such spontaneous activity mechanisms (5–7), persistent activity (8) and connectivity in neural networks (9, 10).

Unraveling neural connectivity, however, is a daunting task; even a small culture with $\approx 10^5$ neurons has several million connections. Electrophysiological approaches combined with microscopy and three-dimensional reconstructions (11, 12) have an enormous capability for identifying the connections between any two neurons, and even all of the connections of a single neuron. However, the identification of the statistical properties of the full connection distribution is beyond current capabilities. Monitoring the development of these connections in embryonic stages is even more ambitious, because it serves to link the stages of growth in the culture with those in the brain (13).

We have recently developed a global bath excitation protocol coupled with graph and percolation concepts (14) to extract many properties of the network of inputs into the neurons (9, 10). The percolation approach in our case deals with changes in the connections of the neuronal network, which can be gradually weakened by means of chemical application. Above a critical connectivity there is a continuum of connected neurons that spans a significant fraction of the network, called the *giant connected component*, which gradually grows larger as the connectivity increases. Below the critical connectivity, the connected neurons are confined in isolated clusters that reduce in size as the connectivity decreases.

In our previous work (9, 10) we were able to follow the disintegration of the giant component as the synaptic strength between neurons was lowered. We observed that the connectivity exhibited a percolation transition as the giant component disintegrated, and characterized the transition by a power law with an effective exponent $\beta \simeq 0.65$. Together with a bond–percolation model on a graph we showed that this exponent is consistent with a Gaussian degree distribution of the input connections.

In this work we extend the percolation approach to quantify connectivity in neural cultures. We first follow developmental aspects of the network, for instance, how the maturing synapses influence connectivity and at which moment a giant connected component emerges. We next derive quantitative information about the connectivity of the network, and extract the average number of connections per neuron and the ratio between excitation and inhibition. Our approach allows us to uncover relevant aspects of the structure of the network that are in general very difficult to extract by using physiological techniques.

## Materials and Methods

The methodology of the percolation approach relies on monitoring the activity of a subsample of the network after a global stimulation given to the entire network. The subsample presently covers $\approx$600 neurons, and although this is only $\approx$1% of the whole network, it is much larger than what any other measuring technique currently offers and can be enlarged if needed. The activity that is measured locally is sensitive to the global connectivity of the network, that is, also to the connections that come from neurons that are not monitored directly. This allows us to uncover the fundamental aspects of the connectivity of the whole network, in particular, aspects such as the average number of connections per neuron ($\bar{k}$), the distribution of these connections ($p_k$), and the size of the largest connected cluster of neurons ($g$).

Two ingredients characterize the percolation transition. The change in connectivity is governed by a *control parameter*, which in turn can be fixed either by varying the synaptic strength between neurons or the number of days in development of the culture. The number of neurons that respond to excitation is the *order parameter*, which measures changes in the network. A sudden jump in the order parameter indicates the appearance of a single connected cluster, and the largest cluster in our subsample is associated with a percolating cluster called the *giant component* (9). The value of the control parameter at which the transition occurs is special, termed a *critical point*. The critical point depends on the connectivity of the network, and therefore reveals information on the structure of the network that is often extremely hard to extract otherwise.

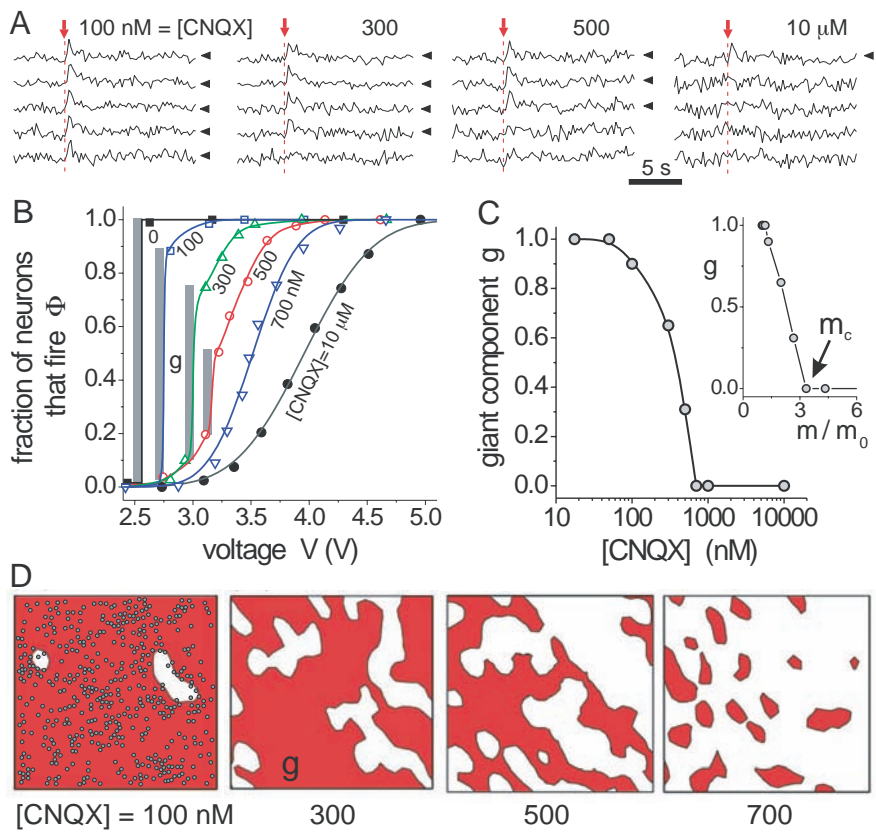

Fig. 1. Network's response and giant component. (*A*) Example of the fluorescence signal of five neurons of a $G_{EI}$ hippocampal network for four concentrations of CNQX and $V = 3.1$ V. Vertical arrows indicate the excitation time, and arrow tips the responding neurons. (*B*) Corresponding response curves $\Phi(V)$ for a total of 450 neurons. Gray bars show the size of the giant component. Lines are a guide for the eye except for 700 nM and 10 μM, which are fits to error functions. (*C*) Corresponding size of the giant component $g$ as a function of [CNQX] (main plot) and as a function of the control parameter $m/m_0$ that quantifies the average connectivity of the network (*Inset*), defined in *Materials and Methods*. (*D*) Spatial coverage of the giant component (red) for the response curves shown in *B*. Dark circles in *Left* are neurons.

**Monitoring Network Activity and Pharmacology.** Neuronal cultures [see supporting information (SI) Fig. S1], grown on 13-mm glass cover slips, were placed in a chamber mounted on a Zeiss inverted microscope with a 10× objective. The neurons were electrically stimulated by applying a 20-ms bipolar pulse through bath electrodes that run along opposite sides of the culture, delivered by a computer-controlled current source, and the corresponding voltage drop $V$ was measured with an oscilloscope (9). Images of calcium-sensitive fluorescence were captured with a cooled CCD camera at a rate of 5 frames per second, and processed to record the fluorescence intensity of 400–600 individual neurons in a region of $830 \times 670$ μm² as a function of time (Fig. 1*A*). Experiments were carried out at room temperature. (See *SI Text* and Fig. S2 and S3 for additional details.)

The network was weakened by gradually blocking the α-amino-3-hydroxy-5-methyl-4-isoxazolepropionic acid (AMPA) glutamate receptors of excitatory synapses with increasing amounts of 6-cyano-7-nitroquinoxaline-2,3-dione (CNQX). *N*-Methyl-D-aspartate (NMDA) receptors were completely blocked with 20 μM of the corresponding antagonist 2-amino-5-phosphonovalerate (APV) so that the disintegration of the network is due solely to CNQX.

To study the role of inhibition, inhibitory synapses were left either active or blocked with 40 μM GABA$_A$ of the receptor antagonist bicuculine. To study the disintegration of the network with and without inhibition, we label the network containing both excitatory and inhibitory synapses by $G_{EI}$, and the network with excitatory synapses only by $G_E$.

**Quantifying Connectivity: Giant Component.** The network's response to a given CNQX concentration was measured as the fraction of neurons $\Phi$ that responded to the electric stimulation at voltage $V$ (Fig. 1*B*), as described in ref. 9. At one extreme, a fully connected network ([CNQX] = 0) leads to a very sharp response curve, because a small number of responding neurons suffice to activate the entire network. At the other extreme, with high concentrations of CNQX ($\simeq 10$ μM), the network is completely disconnected and the response curve is given by the individual neuron's response. $\Phi(V)$ for independent neurons is then well described by an error function (9). For intermediate concentrations of antagonist some of the neurons break off into separated clusters, but a giant cluster still contains a finite fraction of the network.

The size of the giant component $g$ is measured as the biggest fraction of neurons that fire together in response to the external excitation (Fig. 1*B*). The size of the giant component decreases with the concentration of antagonist (Fig. 1*C*), and it is considered to be zero when a characteristic jump is not identifiable. Conceptually, the presence of a giant component reveals the existence of long-range connectivity that spans the entire network. Fig. 1*D* shows the spatial coverage of the giant component (within the field of view of the microscope) during the disintegration of the network for the response curves of Fig. 1*B*. For [CNQX] = 0 the giant component comprises the entire network. As the concentration of CNQX increases the giant component reduces in size, but it covers a continuous area that extends the entire network. At a critical concentration, [CNQX] $\simeq 700$ nM, a giant component is not identifiable and the group of neurons that fire together in response to the excitation correspond to isolated clusters.

**Characterization of the Control Parameter.** To quantify the change in connectivity of the network as it disintegrates we introduce a control parameter that measures the average number of inputs $m$ required for a neuron to fire, and provide an expression that relates $m$ with the concentration of CNQX.

Our model assumes that each input onto a neuron increases or decreases its threshold voltage $V_T$, depending on the polarizing nature of the input, excitatory or inhibitory. All input connections are assumed to have the same synaptic efficacy $g_{syn}$, all inputs are synchronous, and we assume that the synaptic strength between neurons is not affected by the electric stimulation (see also *SI Text, Experimental Procedures*). Each input has a contribution $g_{syn}$ to the total voltage, and hence $m = V_T/g_{syn}$ inputs are required to excite a neuron and pass the signal in the network.

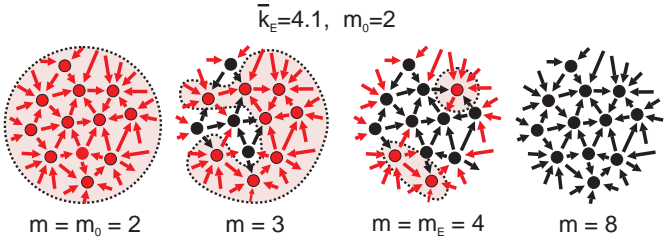

**Fig. 2.** Control parameter $m$ and giant component. Schematic disintegration of a network with only excitatory inputs, average connectivity $\bar{k}_E = 4.1$ and $m_0 = 2$. $m$ increases as the synaptic coupling between neurons is reduced. Neurons (circles) having at least $m$ inputs (arrows) fire and pass the signal on (red); the rest remain inactive (black). A giant cluster (outlined areas) initially connects most of the network and decreases in size as more neurons become inactive. At a critical value $m = m_E \simeq \bar{k}_E$ the network disintegrates and the giant component breaks off into isolated clusters.

The application of CNQX gradually reduces the synaptic strength between neurons and effectively decreases $g_{syn}$, so that the number of inputs $m$ required for a neuron to fire gradually increases. To obtain a relation between $m$ and the concentration of CNQX, we first observe that the synaptic strength between neurons can be quantified as the fraction $c$ of receptor molecules that are not bound to the antagonist CNQX and therefore are free to activate the synapse. This fraction is given by $c = 1/(1 + [\text{CNQX}]/K_d)$, with $K_d = 300$ nM (9) (*SI Text*, Fig. S3), and takes values between 0 (full blocking) and 1 (full connectivity). Thus, $g_{syn}$ decreases to $c\, g_{syn}$ as the concentration of CNQX increases. The effective number of inputs necessary to excite a neuron can be then written as $m = V_T/g_{syn} = m_0/c$, where $m_0$ is the number of inputs required for a neuron to fire in the unperturbed network.

Physiological studies provide values of the threshold voltage $V_T$ on the order of 30 mV (24). There is considerable variability in the measured values for $g_{syn}$ (see Table S1) but because we are averaging over many neurons we are sensitive to the average value, which is much more reliably measured as 2 mV. In consequence, approximately $m_0 = 15$ inputs are typically required to excite a neuron. Reported values for $m_0$ are in the range of 5–30 (25, 26). Hence, although $m = m_0/c$ is the natural variable to quantify the change in connectivity in the network, because of the uncertainty in $m_0$, we define our final order parameter in the form of $m/m_0 = 1 + [\text{CNQX}]/K_d$.

The dependence of the giant component on $m/m_0$ for a particular experiment is shown in Fig. 1C. The giant component gradually reduces in size as $m/m_0$ increases. Above a normalized critical value denoted $m_c = m/m_0$, the giant component disintegrates.

**Giant Component, Connectivity, and Amount of Inhibition in the Network.** The disintegration process of the giant component in terms of $m$ is illustrated in Fig. 2. Conceptually, $m$ quantifies the average number of inputs that a neuron needs to fire. For the unperturbed network we have $m = m_0 \ll \bar{k}$, where $\bar{k}$ is the average connectivity of the network. Hence, all neurons fire and the giant component comprises of the entire network. As the connectivity decreases and $m$ grows, those neurons having less than $m$ inputs get disconnected from the network and, in turn, reduce the number of inputs on their target neurons. The size of the giant component gradually decreases. At a critical value of $m$ (denoted by $m_E$) the giant component disintegrates and the network is comprised of isolated clusters.

The point $m = m_E$ characterizes the critical point of the percolation transition. For $m < m_E$ the network is connected through a giant component. For $m > m_E$ the network comprises isolated clusters, or single neurons at the extreme of full blocking of the network. In our experiments we consider networks with excitatory and inhibitory inputs ($G_{EI}$) and networks with excitatory inputs only ($G_E$). Therefore, we will measure two critical points, and label them $m_{EI}$ and $m_E$, respectively.

The precise values of $m_E$ and $m_{EI}$ depend on the distribution of connections in the network, $p_k(k)$. In the *SI Text*, we introduce a percolation model to study the disintegration of the network in terms of $m$ for different $p_k(k)$. We particularly treat the case where $p_k(k)$ corresponds to a Poisson-like distribution, which describes well the connectivity in neural cultures (9), and show that $m_E \simeq \bar{k}_E$, where $\bar{k}_E$ is the average number of input connections per neuron. Thus, $m_E$ provides a direct estimation of the average number of excitatory inputs per neuron.

Inhibition effectively reduces the average connectivity of the network (9) and therefore the difference between the values of $m_{EI}$ ($G_{EI}$) and $m_E$ ($G_E$ networks) contains information on the amount of inhibition in the network. If $\bar{k}_E$ and $\bar{k}_I$ denote the average number of excitatory and inhibitory inputs per neuron, respectively, then $m_{EI} \simeq \bar{k}_E - \bar{k}_I$ and $m_E \simeq \bar{k}_E$. Hence, the ratio between inhibition and excitation in the network is given by $\bar{k}_I/\bar{k}_E = 1 - m_{EI}/m_E$ (see *SI Text* for details).

Finally, because of the uncertainty in the value of $m_0$ discussed above, we use the critical points normalized in units of $m_0$ in the analysis of the experimental data, and label them $m_c \equiv m_{EI}/m_0$ and $m'_c \equiv m_E/m_0$, respectively.

**Measuring the Development of the Network.** The developmental time of the network (culture days *in vitro*, DIV) can be viewed as playing a role opposite to the CNQX concentration described above. Neurons that are initially isolated start to connect to each other as they develop processes. At a critical time $t_c$ a giant component emerges, connecting the entire network.

Experiments were carried out with hippocampal cultures derived from embryonic brains with either 17 (E17) or 19 (E19) days of development, and postnatal brains just after birth (P0). To monitor the development of connections in the network we considered two scheduling protocols. In the first approach we measured the response curve $\Phi(V)$ of cultures from a batch of identical cultures at 4-h intervals. The size of the giant component $g$ was analyzed at every step, and the process repeated until $g = 1$ was attained, typically over 48 h. The critical time $t_c$ of the emergence of the giant component was obtained by fitting the curves $g(t)$ to a power law of the form $g(t) \approx |1 - t/t_c|^\gamma$.

In the second protocol we prepared a batch containing 24 identical E17 hippocampal cultures, and measured the response curves $\Phi(V)$ at 1-day intervals for 12 days. We used one-half of the batch to study the disintegration of the $G_{EI}$ network and the other half to study the $G_E$ one. From the analysis of the response curves we obtained the evolution of the critical points $m_c$ and $m'_c$ as a function of developmental time.

## Results

**Network Development.** Examples of neurons' responses during development are shown in Fig. 3A. Neurons have a weak response to the external excitation at the early stages of development. The response becomes stronger at later stages and spontaneous activity starts to emerge. The development of connectivity is monitored through the evolution of the response curves $\Phi(V)$, as shown in Fig. 3B. During the first 2 days *in vitro* (DIV) of E19 neurons there is no response to an external excitation. At day 2 excitability is established, but it necessitates relatively high voltages. As the culture matures neurons become progressively more sensitive and the excitation threshold decreases. During the early developmental stage of days 2–3 the response curves are well described by error functions (see *Materials and Methods*), as shown in Fig. 3*B*, indicating that the neurons are either disconnected or forming isolated clusters.

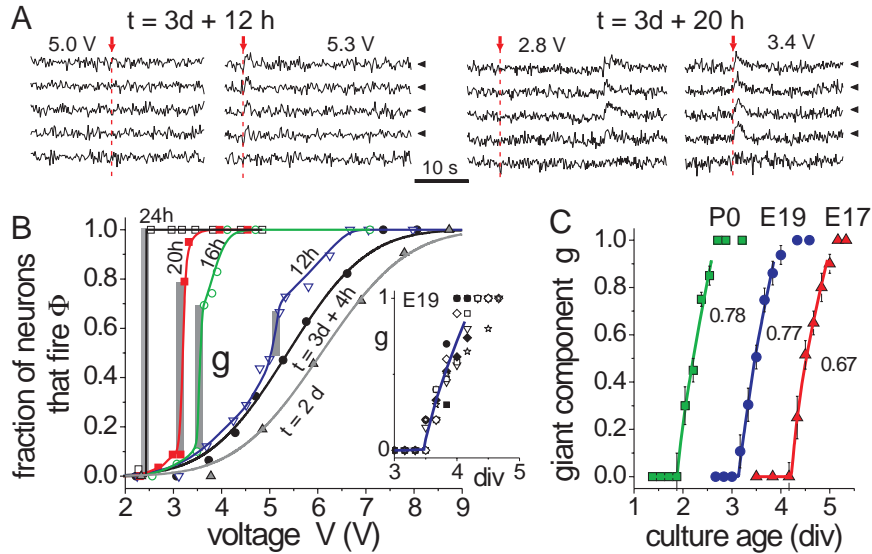

**Fig. 3.** Development of the network. (*A*) Fluorescence signal of five E19 hippocampal neurons at different developmental stages and voltages. Vertical arrows indicate the excitation time, and arrow tips the responding neurons. (*B*) Corresponding response curves Φ(*V*). The gray bars show the size of the giant component. Lines are a guide for the eye except for $t = 2$ days and $t = 3$ days $+ 4$ h that correspond to fits to error functions. (*Inset*) Size of the giant component as a function of time for seven experiments with E19 cultures. The line is a power law fit of the averaged data. (*C*) Size of the giant component as a function of time for hippocampal cultures derived from brains at three different developmental stages: E17 (2 experiments), E19 (7 experiments), and P0 (2 experiments). Lines are power law fits. The values indicate the critical exponent $\gamma$.

At approximately day 3 the response curves change qualitatively. As the voltage is ramped up a small jump appears that signals a group of neurons that always fire together, and are therefore connected. This characterizes the emergence of a giant connected component (see *Materials and Methods*), whose size *g* grows rapidly in the following hours until it encompasses the whole network. The whole network fires together when $g = 1$, which is attained at approximately day 4. Although the network can obviously evolve further, the size of the giant component does not.

The size of the giant component *g* changes with the culture age (days *in vitro*) in a reproducible fashion, as shown in Fig. 3*B Inset* for seven experiments using E19 cells. The averaged data are shown in Fig. 3*C*. We see that the giant component emerges at $t_c \simeq 3.1$ days and grows in time with a behavior that is well described by a power law with exponent $\gamma \simeq 0.77$. Within just over a day the giant component is created.

Because E19 cultures take ≈2 days to start their electrical activity, and slightly more than another day to develop the giant component, the actual time at which the giant component matures coincides with the full term of the pregnancy of the rat. To check whether this is a general situation we looked at the development of the same process in neurons derived from 17-day-old embryos (E17) and newborn postnatal pups (P0).

As shown in Fig. 3*C*, the evolution of the giant component is very similar for both cases, but it is delayed by the different time during which the electrical activity of the neurons begins. For E17 cultures a response to external electrical stimulation appears only at day 3. The giant component emerges at $t_c \simeq 4.2$ days, with $\gamma \simeq 0.67$, and the giant component is fully developed by day 5. On the contrary, for P0 cultures the development start earlier. Neurons respond to the external excitation 1 day after plating. The giant component appears at $t_c \simeq 1.9$ days, with $\gamma \simeq 0.78$.

Thus, for both E17 and E19 the day at which the giant component matures is very close to the day of full term.

Overall, the development of the giant component is similar for the three culture age types we examined. The existence of a critical transition from no connected component to one that is rapidly growing with time repeats itself. The effective power law growth rate of the giant component is similar for all three culture types and is ≈0.7 within an error of ≈10%.

We also investigated the generation of spontaneous activity during development (see *SI Text* and Fig. S4). We identified a group of neurons that tended to fire together simultaneously in a spontaneous manner and took the largest such fraction of neurons as a measure of the level of spontaneous activity. We observed that the level of activity increased at the same rate as the size of the giant component, and that the occurrence of spontaneous activity extending the entire culture coincided with $g = 1$, full connectivity of the network.

**Average Connectivity of the Network.** As described in *Materials and Methods*, the definition of the control parameter in terms of the number of inputs *m* that are needed for excitation yields a coarse measure of the average input connectivity $\bar{k}$ in the network. Although for bond percolation models, where links are broken rather than weakened, this would be an exact result, here the result is approximate. However, when we measure *relative* changes, the results can be considerably more accurate.

The behavior of the giant component as a function of *m* for hippocampal and cortical cultures is shown in Fig. 4*A*. For small values of *m* the network remains fully connected with $g = 1$, and the effect of reducing the synaptic coupling is observed only in the decay of spontaneous activity (data not shown). At higher values of *m* the network starts to disintegrate and the giant component reduces in size. For networks with excitatory and inhibitory inputs ($G_{EI}$) the giant component reaches zero at the normalized critical value $m_c$. The removal of the inhibitory component of the network by the administration of bicuculine ($G_E$ networks) leads to an effective displacement of the entire curve to higher *m*, because less excitatory inputs are necessary when the offset from the inhibitory inputs is absent. Therefore, $G_E$ networks always disintegrate at a critical value $m'_c > m_c$.

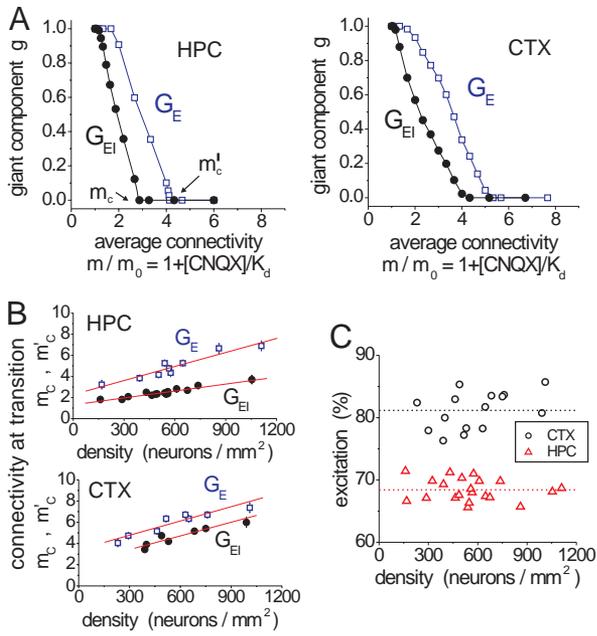

**Fig. 4.** Average connectivity of the network and amount of excitation. (*A*) Size of the giant component as a function of $m/m_0$. (*Left*) E19 hippocampal (HPC) cultures (black, $G_{EI}$ networks, 16 experiments; blue, $G_E$ networks, 9 experiments). (*Right*) E19 cortical (CTX) cultures (black, $G_{EI}$, 7 experiments; blue, $G_E$, 8 experiments). Lines are a guide for the eye. (*B*) Variation of the critical points $m_c$ and $m'_c$ with the density of the neural culture, for $G_E$ (squares) and $G_{EI}$ (dots). (*Upper*) E19 hippocampal cultures. (*Lower*) E19 cortical cultures. Lines are least-squares fits. (*C*) Percentage of excitatory inputs as a function of the density for E19 hippocampal (triangles) and cortical (circles) cultures. Lines correspond to the average value.

The disintegration of the network for both hippocampal and cortical cultures shows the same characteristic behavior. Cortical cultures, however, have critical values that are larger by ≈25% than for hippocampal ones. This indicates that the average connectivity is larger by the same ratio for cortical cultures.

It is interesting to ascertain whether the average input connectivity of the network varies with the density of the neural culture. Although Fig. 4*A* is an average over several datasets, in Fig. 4*B* we plot in detail the critical points for each culture as a function of its density. We see that the critical points $m_c$ and $m'_c$ grow linearly with the density. The values for cortical cultures are higher than for hippocampal ones, again indicating that cortical cultures have ≈25% more input connections than hippocampal ones.

As explained in *Materials and Methods*, $m'_c$ is a measure of the average number of excitatory inputs in the network, $\bar{k}_E$. The range of variation of $m'_c$ shows that hippocampal cultures have on the order of $\bar{k}_E \approx 3\, m_0$ excitatory inputs per neuron at the lowest density (150 neurons per mm²), and this grows by a factor 2 when the density increases by a factor 8. For cortical cultures $\bar{k}_E \approx 4\, m_0$ for the lowest density, growing again by a factor 2 at the highest densities studied.

**Identifying the Inhibitory Component of the Network.** The relative difference between the two critical values for appearance of the giant component can be used to quantify the amount of inhibition in the network. The ratio between the average number of inhibitory $\bar{k}_I$ and excitatory $\bar{k}_E$ inputs in the network is then given by $\bar{k}_I/\bar{k}_E = 1 - m_c/m'_c$ (see *Materials and Methods*).

As Fig. 4*B* shows, in the whole range of densities studied, there is a different behavior of the two critical points. The hippocam-pal and cortical cultures also differ, because the lines of critical points $m_c$ and $m'_c$ are closer for cortical cultures, indicating that the amount of excitation is higher. A quantitative analysis for both cultures types is shown in Fig. 4*C*. As expected, the amount of excitation is independent of the density and does depend on the culture type. Hippocampal cultures are found to have ≈70% excitation whereas cortical cultures have ≈80%. These results are consistent with physiological studies (5, 15, 16) which reported that hippocampal cultures have ≈70% excitation, an amount that increases to ≈85% for cortical cultures.

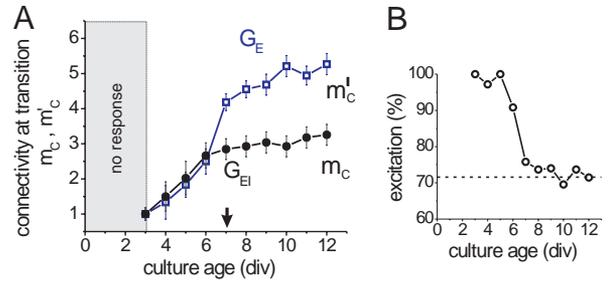

**Fig. 5.** Emergence of inhibition during development. (*A*) Evolution of the critical points $m_c$ ($G_{EI}$) and $m'_c$ ($G_E$ networks) as a function of time for E17 hippocampal cultures, averaged over two experiments. The arrow indicates the time of emergence of inhibition. (*B*) Corresponding evolution of the amount of excitation. The dashed line shows the amount of excitation averaged over days 10–12.

**Emergence of Inhibition During Development.** During the early stages of development of neural networks all neurons are excitatory. GABA, the main inhibitory neurotransmitter, has a depolarizing action and therefore has an excitatory role (2). This role disappears at later stages of development and inhibition emerges.

Our analysis of the critical points $m_c$ and $m'_c$ can be used to identify the stage of development at which the inhibitory switch takes place. To map out this process we have used E17 cultures and measured the critical points $m_c$ and $m'_c$ at 24 h intervals for ≈2 weeks (see *Materials and Methods*).

The variation of the critical points $m_c$ and $m'_c$ as a function of the developmental time *in vitro* is shown in Fig. 5*A*. During the first 3 days after plating the neurons do not respond to the external excitation. Response to stimulation appears at DIV 3–4, and a giant component starts to be identifiable.

As the network matures and the average connectivity increases, disintegration of the giant component occurs at larger concentrations of CNQX, and hence, the values of $m_c$ and $m'_c$ gradually increase. The two critical points grow similarly during the first days of development, which indicates that all neurons are excitatory. It is not until day 7 that we clearly measure different values. This signals the emergence of inhibition.

The critical point $m'_c$ increases during the next 4 days as more connections emerge in the network. The rate of increase of $m_c$, however, is significantly lower and indicates that the number of inhibitory neurons gradually increases. As shown in Fig. 5*B*, the average relative fraction of excitatory inputs per neuron rapidly decreases in the network after day 6, and reaches a constant value of ≈70% excitatory inputs at day 10–12. This corresponds to full maturation of the network, and hence the amount of excitatory inputs that we measure is similar to the one obtained for mature E19 hippocampal cultures (Fig. 4*C*).

## Discussion

During the first days *in vitro* the neural network exhibited no activity (DIV 1 for P0, 1–2 for E19, and 1–3 for E17), and the gradual emergence of spontaneous bursts occurred during the next few days. This development of excitability may be the result of several factors, related to both single-neuron and whole-network characteristics. The changes we observe in the threshold for activation of the network (Fig. 3A) indicate that the threshold voltage $V_T$ is decreasing with time, whereas maturation of synapses may affect the average electric input from each synapse $g_{syn}$. The emergence and extension of axons from cell somata that take place during development (16) will affect the probability distribution function of connections in the network $p_k$. As a result, during development we can identify the emergence of large-scale connectivity associated with the giant connected component, but cannot pinpoint the average number of connections. In contrast, observing the gradual breakdown of connectivity in the network because of increases in the CNQX concentration gives the ability to quantify $m = V_T/g_{syn}$. This is because at a given measurement time we hold factors, such as the threshold $V_T$ and connection distribution $p_k$ constant, and by varying the CNQX concentration we are only changing $g_{syn}$ (and therefore $m$) in a well defined manner.

The time $t_c$ at which the giant component emerges changes by $\approx 1.2$ days from E17 to E19 cultures, and by the same amount to P0. This indicates that the timing of the emergence of the giant component is shifted according to the development of the embryonic brains. In E17 and E19 cultures the giant component reaches $g = 1$ at DIV $5 \pm 0.5$ and $4 \pm 0.5$, respectively, whereas in P0 the network has a giant component almost immediately when it becomes active. Taking into account the uncertainty in the precise conception time of the rat ($\approx 0.7$ day) and the recovery of the neurons after plating, the cultures all achieve full connectivity by E21–E22, about the expected time of birth. We therefore hypothesize that, during the development of the rat brain, a state of full connectivity is achieved close to delivery. The importance of the day of birth for the maturation of neural circuits has been observed by others in a variety of circumstances (18).

The observation that denser cultures have higher connectivity is an interesting one. The linear relation of density to inputs can explain some of the observations regarding burst initiation in cultures. In one-dimensional cultures, we have shown that Burst Initiation Zones (BIZs) are preferably localized to high-density areas (19). We therefore conjecture that burst initiation can be ascribed to a subset of neurons that have more inputs, and are therefore more sensitive to background activity (20).

The emergence of inhibition takes place at day 7–8 for E17 hippocampal cultures. This result is in agreement with the study of Ganguly et al. (21) with E18 hippocampal cultures, where the GABA switch takes places at day 8–9. They also observed that the GABA switch was gradual and that it required $\approx 4–5$ days to complete, in agreement with our results.

The quantitative precision of our analysis is limited by the assumption that the critical points $m_c = m_{EI}/m_0$ and $m_c' = m_E/m_0$ vary linearly with the average number of excitatory and inhibitory inputs, $\bar{k}_E$ and $\bar{k}_I$, respectively. In the *SI Text* we treat this problem and show numerically that the relation of $m_{EI}$ and $m_E$ with $\bar{k}_E$ and $\bar{k}_I$ is indeed linear, with a slope in the range 1–2, depending on the details of the distribution of the input connections. Because we have measured this distribution in our networks to be Gaussian or Poisson-like (9), we take the slope to be 1.3, corresponding to such a distribution (*SI Text*, Fig. S5). Using these assumptions we obtain, as a function of density and for $m_0 \simeq 15$, values of 60–120 and 75–150 for the average number of excitatory inputs per neuron in hippocampal and cortical cultures, respectively. The possible error is therefore determined by the range of variation in $m_0$ reported in the literature (see Table S1), which is between 5 and 30. This gives an overall possible error of a factor 0.33–2, and a final spread of possible values in the range of 20–120 inputs in the low-density hippocampal cultures. A better determination of $m_0$ will certainly improve the confidence in the number of input connections measured. These values are about two orders of magnitude lower than the average connectivity in the brain, which is in the range $1–8 \times 10^3$ synaptic connections per neuron on average (22, 23). This may have to do not only with the ability of neurons in the brain to use the third dimension for connections, but also with the continuous refinement of connectivity based on sensory experience.

**ACKNOWLEDGMENTS.** We thank M. Segal, J.-P. Eckmann, O. Feinerman, S. Jacobi, and A. Rotem for fruitful discussions. This work was supported by Israel Science Foundation Grant 993/05 and by the Minerva Foundation, Germany. J. S. was supported by EU Training Network PHYNECS Grant HPRN-CT-2002-00312 and by the Curwen-Lowy Fellowship.